\documentclass[10pt,notitlepage,pra,twocolumn]{revtex4-1}
\usepackage{amsfonts}
\usepackage{amsmath}
\usepackage{amssymb}
\usepackage{graphicx}
\usepackage{float}
\usepackage[toc,page,header]{appendix}
\usepackage{color}

\begin{document}

\title{Dynamical phases in a one-dimensional chain of Heterospecies Rydberg atoms with next-nearest neighbor interactions}
\author{Jing Qian$^\dagger$, Lu Zhang, Jingjing Zhai and Weiping Zhang}
\affiliation{Quantum Institute for Light and Atoms, Department of Physics, East China
Normal University, Shanghai 200062, People's Republic of China}

\begin{abstract}
We theoretically investigate the dynamical phase diagram of a one-dimensional chain of laser-excited two-species Rydberg atoms. The existence of a variety of unique dynamical phases in the experimentally-achievable parameter region is predicted under the mean-field approximation, and the change of those phases when the effect of the next-nearest neighbor interaction is included is further discussed. In particular we find the competition of the strong Rydberg-Rydberg interactions and the optical excitation imbalance can lead to the presence of complex multiple chaotic phases, which are highly sensitive to the initial Rydberg-state population and the strength of the next-nearest neighbor interactions.

\end{abstract}

\maketitle
\preprint{}

\section{Introduction}
Ultracold atoms combined with optical lattices \cite{jaksch98,bloch05,morsch06,lewenstein07,lewenstein12} can give rise to a clean and controllable platform for simulating and studying quantum many-body physics \cite{bloch08}, capably demonstrating various quantum phases and even the dynamics of phase transitions. 
Recently, an efficient optical-lattice trap for Rydberg atoms has been realized in the experiment \cite{anderson11}, extending quantum simulation researches to the realm of Rydberg atom physics.
Different from the common-used ground-state atoms, the giant dipole moment induced by the highly excited atomic states results in an interaction of unprecedented magnitude and range between two Rydberg atoms, that is, the Rydberg-Rydberg interactions (RRIs) \cite{gallagher08}, which could further lead to the Rydberg blockade effect \cite{comparat10}. This character of Rydberg atoms is conducive to simulate the strong-correlated quantum many-body system \cite{pohl10,weimer10,viteau11,schaub12} and to realize non-equilibrium quantum phase transitions \cite{carr13}.

%In combination with optical lattices \cite{bloch05,morsch06,lewenstein07,lewenstein12} and Feshbach resonance \cite{chin10,forges15}, ultracold atoms can give rise to a clean and controllable platform for studying quantum many-body systems \cite{bloch08}, capably demonstrating various dynamical phases and even the dynamics of phase transitions. 
%A substantial research effort has initially been undertaken on simulating Mott-superfluid phase transition with ground state single-species atoms \cite{jaksch98,greiner02,xu14}. 
%With atoms of two different species included, the wealth of phases grows to a daunting complexity \cite{bhaseen09}.
%Quite a few novel phases and phase separations emerge due to the coexistence of both interspecies and intraspecies interactions \cite{mishra07,papp08,zhou08,lv14}. 
%
%
%Moreover, different from ground-state atoms, the giant dipolar moment induced by the highly excited atomic states can offer an unprecedented magnitude and range of interaction between Rydberg atoms, that is, the Rydberg-Rydberg interactions (RRIs) \cite{gallagher08,comparat10}. By virtue of these strong interactions, laser excitation of high-energy Rydberg states has enabled numerous breakthroughs for exploring quantum many-body physics in strong-correlated systems \cite{weimer10,schaub12}. Recently an efficient optical-lattice trapping for Rydberg atoms has been realized in experiment \cite{anderson11}. 
%This adds to the possibility of studying strong-correlated quantum phases with a Rydberg lattice gas. 

So far, few attention is paid to the heterospecies Rydberg lattice gas, most researches focus on the dynamical phases of single-species Rydberg lattice gases when both strong RRIs and spontaneous emission effects are present \cite{lee11,qian12,qian13,hoening14,olmos14,chan15}. 
For a single-species Rydberg lattice gas, intrinsically, only the strong RRIs in the nearest-neighbor (NN) sites should be considered \cite{viteau11}, since the long-range RRIs due to next-nearest neighbor (NNN) interactions are already orders of magnitude smaller \cite{honing13}.
However, things will be quite different for the heterospecies case. Here the heterospecies case could be two Rydberg atoms of different atomic species or two same species atoms occupied in different Rydberg hyperfine states \cite{hsueh13}. 
The different excitation frequencies would disrupt the Rydberg blockade mechanism as well as those phases existing in the single-species case. Moreover, the RRIs between heterospecies Rydberg atoms can vary by orders of magnitude \cite{teixeira15} which means the NNN interaction between two Rydberg atoms of same species could be comparable to the NN interaction between heterospecies atoms, and vice versa. So the effect of NNN interaction deserves to be discussed in the lattice model of heterospecies Rydberg atoms. Especially, the importance of NNN interactions has been demonstrated by a recent experiment that the excitation dynamics could be significantly different from the common cases \cite{barredo15}.

In the present work we explore the dynamical phase diagram of a one-dimensional (1D) chain of heterospecies Rydberg atoms under an open environment. The Rydberg atoms of two different species are alternatively arranged in the 1D optical lattice with internal states being subjected to the laser pumping and spontaneous decay. By the mean-field approximation, we predict the presence of a rich variety of dynamical phases, involving three stable phases which are the antiferromagnetic phase, the bistable antiferromagnetic phase, and the tristable antiferromagnetic phase, and unstable phases whose dynamics can change from ordinary oscillation to chaos under strong RRI case. No uniform phase that presents in the single-species case is found due to the heterospecies atomic excitations. 
We investigate the impacts of repulsive and attractive NNN interactions on those phases and find the stable phases prefer the repulsive ones. 
Especially, the chaotic phase has been shown high sensitivity to the NNN interactions, which verifies the necessity of including them in our model of heterospecies Rydberg atom chain.
 
The paper is organized as follows: in section II we present a scheme for presenting the exciting dynamics of a heterospecies Rydberg atom chain and derive basic master equations, from which the stationary state solutions can be solved. In sections III and IV, we investigate the change of phase diagrams without and with the effect of NNN RRIs, respectively. In section V, we compare the different influences of repulsive and attractive NNN RRIs on the chaotic phase. Finally, a brief conclusion is given in section VI.

\section{Scheme and master equation}

\begin{figure}
\includegraphics[width=3.44in,height=1.4in]{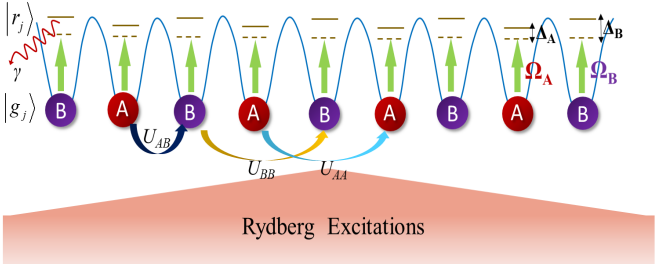}
\caption{(Color online) Schematic of two-species atoms ($A$ and $B$) trapped in a 1D optical lattice. Atoms are excited from the ground state $\left \vert g_j \right \rangle$ to the Rydberg state $\left \vert r_j \right \rangle$ by different laser Rabi frequencies $\Omega_A$ and $\Omega_B$. Simultaneously, they are suffering from non-ignorable spontaneous decay $\gamma$. The NN and NNN RRIs strength are, respectively, labeled by $U_{AB}$ and $U_{AA(BB)}$, representing the interspecies short-distance and intraspecies long-distance interactions. The one-photon detunings between the laser frequency and the atomic transition frequency are denoted by $\Delta_{A}$ and $\Delta_{B}$, differently.}
\label{model}
\end{figure}

As shown in Fig. \ref{model}, the system we propose consists of a chain formed by two-species Rydberg atoms, trapped in a regular 1D lattice. An efficient trapping of single rubidium atoms in a 1D optical lattice was initially realized in \cite{anderson11}, and the loading of different atoms can be using a species-selective optical lattice \cite{scelle13,ramos14}. 
Here we assume the atoms of species $A$ and $B$ are alternatively arranged in the lattice sites with a filling factor of one atom per site and the hopping of an atom into an adjacent filled site is forbidden by the deep depth of the lattice.
Each atom $j$ is modeled as a two-level configuration, composed of a ground state $\left \vert g_{j} \right \rangle$ and a Rydberg state $\left \vert r_{j} \right \rangle$, whose transition is performed by an off-resonant laser beam with the Rabi frequency $\Omega_{j}$ and the one-photon detuning $\Delta_{j}$. For the lattice structure here, if we replace the site index by the atom species inside, i.e., $j=A$, we will have $j\pm 1=B$ and $j\pm 2=A$. 
It is also worth noting that the assumption of two-species atoms can be equivalent to a scheme of same ground-state atoms excited to two different Rydberg hyperfine states \cite{hsueh13}.

In the absence of external fields, two atoms prepared in the $nS$-Rydberg states generally interact via a non-resonant van der Waals (vdWs) RRI, described by $U_{ij}=C_6^{(ij)}/|x_i-x_j|^6$, where $x_{i(j)}$ represents the position of atom $i(j)$ in the lattice and $C_6^{(ij)}$ the coefficient for dispersion. To our knowledge, $C_6^{(ij)}$ is well-defined \cite{singer05} and measured \cite{beguin13} in the case of same species of atoms, such as Rb-Rb, Na-Na, K-K, Li-Li, Cs-Cs. However, as for different atomic species or different hyperfine states,
$C_6^{(ij)}$ changes significantly \cite{olmos11,zoubi15}. 
In the present, we focus on two vdWs-type interactions: 
i) The interspecies interaction between the NN sites of different atomic species, denoted by $U_{j,j\pm 1}=U_{AB}$; 
ii) The intraspecies interaction between the NNN sites of same atomic species, denoted by $U_{j,j\pm2}=U_{AA}$ or $U_{BB}$.

Here we consider both the NN and NNN interactions, since the long-range NNN RRIs between same species of atoms are no longer negligible once the intraspecies interactions are much stronger than the interspecies ones. Other longer-range interactions, such as the next-next-nearest neighbor interaction, is at least by a factor  $(C_6^{(AB)}/C_6^{(AA(BB))})(2/3)^6$ smaller than the NNN interactions, and therefore are negligible here. Then in the frame rotating with the laser frequency, the Hamiltonian of the system reads
\begin{equation}
	\mathcal{H} = \sum_{j} ( {\mathcal{H}_{j}}+\sum_{k=j\pm 1,j\pm 2} U_{jk}\left \vert r_j \right\rangle \left \langle r_j \right \vert \otimes \left \vert r_k \right\rangle \left \langle r_k \right \vert ),
	\label{Hamtot}
\end{equation}
where $\mathcal{H}_{j} =   -\Delta_{j} \left| r_j \right\rangle {{\left\langle r_j \right|}} +{ \Omega_j \left( {\left| g_j \right\rangle {{\left\langle r_j \right|}} + \left\vert r_j\right\rangle\left\langle g_j \right\vert} \right)} $ accounting for the single atom-laser coupling and $U_{jk}=U_{AB}$ for the NN RRIs with $k=j\pm1$; and $U_{jk}=U_{AA}$ or $U_{BB}$ for the NNN RRIs with $k=j\pm2$. The system dynamics is described by the master equation of the  density matrix operator $\rho$ \cite{diehl08}:
%Additionally the spontaneous decay $\gamma$ from the unstable state $\left \vert r_j \right\rangle$ will be included through Lindblad terms, giving rise to the master equation for the density matrix operator $\rho$ \cite{diehl08}:
\begin{equation}
	{\partial _t}\rho  =  - i\left[ {\mathcal{H},\rho } \right] + \mathcal{L}\left[ \rho  \right],
	\label{master}
\end{equation}
where the effect of the spontaneous decay from the unstable state $\left \vert r_j \right\rangle$ with the rate $\gamma$ is included by the Lindblad operator
\begin{equation}
	\mathcal{L}\left[ \rho  \right] = \gamma \sum\limits_j {\left( { - \frac{1}{2}\left\{ {\left| {{r_j}} \right\rangle \left\langle {{r_j}} \right|,\rho } \right\} + \left| {{g_j}} \right\rangle \left\langle {{r_j}} \right|\rho \left| {{r_j}} \right\rangle \left\langle {{g_j}} \right|} \right)} .
\end{equation}

Due to the enormous Hilbert space, it is hard to perform an exact numerical simulation on the above chain model of large atom number, 
so we apply the mean-field approximation (MFA) here. Compared with the method of Monte Carlo simulations \cite{hoening14}, as the interatomic quantum correlation and its fluctuations ignored, the MFA may be failed to predict the phase transition under the same system parameters, or to obtain the exact boundary in the phase diagram. 
However, the MFA is still regarded as a reliable and adequate tool to qualitatively describe the phase diagram and at least to predict the existence of kinds of steady-state phases \cite{lee11,qian12,qian13}. 
By the MFA we can neglect the intersite quantum correlation and factorize the density matrix into each site $\rho=\otimes_j \rho_j$ \cite{diehl10,tomadin11}. For atom $j$, the second term in Eq. (\ref{Hamtot}) should be replaced by $\left| {{r_j}} \right\rangle \left\langle {{r_j}} \right|\sum_{k = j \pm 1,j \pm 2} {{\rho _{k,rr}}} $, where $\rho_{k,ab=gg,gr,rg,rr}$ represent the density matrix elements for the two-level atom in site $k$. Furthermore,  we consider two sublattices filled with atoms $A$ and $B$, respectively, as shown in Fig. \ref{model}. The excitation probabilities of their Rydberg states are different. That is why the uniform phase can not be found here.
We assume, the NNN interactions between same-species atoms are also relatively weak so its induced blockade effect is ignored. 

With all the approximations and the assumptions above, the motional equations of the density matrix can be derived as
\begin{eqnarray}
{{\dot \rho }_{A,rr}} &=& 2\Omega _A\rho_{A,gr}^I - {\rho _{A,rr}}, \label{Arr}\\
{{\dot \rho}_{A,gr}} &=&  i\Delta_{A,eff}{\rho_{A,gr}} +i\Omega_{A} \left( {1-2{\rho _{A,rr}} } \right) -\frac{1 }{2}{\rho_{A,gr}}, \label{Agr}\\
{{\dot \rho }_{B,rr}} &=& 2\Omega_B \rho_{B,gr}^I - {\rho _{B,rr}}, \label{Brr}\\
{{\dot \rho}_{B,gr}} &=&  i\Delta_{B,eff}{\rho_{B,gr}} +i\Omega_{B} \left( 1-2{\rho_{B,rr}} \right) -\frac{1 }{2}{\rho_{B,gr}}, \label{Bgr}
\end{eqnarray}
where all the frequencies are scaled by decay rate $\gamma$, and the effective detunings are defined as
\begin{eqnarray}
	\Delta_{A,eff}=\Delta_{A}-U_{AA}\rho_{A,rr}-U_{AB}\rho_{B,rr}, \label{DeltaA} \\
	\Delta_{B,eff}=\Delta_{B}-U_{BB}\rho_{B,rr}-U_{AB}\rho_{A,rr},\label{DeltaB}
\end{eqnarray}
from which we can find the bare detunings are shifted by two nonlinear terms being proportionial to the NN and NNN interactions, respectively.

In principle, via an adjustment of bare detunings $\Delta_{A}$ and $\Delta_{B}$, one can compensate the density-dependent frequency shifts caused by RRIs so that the effective detunings may vanish. That is the internal working of the anti-blockade effect in the Rydberg chain of single-species atoms \cite{ates07,amthor10}. 
However, both the intraspecies and interspecies RRIs work here, which induce a complicated nonlinear coupling, making that compensation effect become elusive.

We begin our discussions about the dynamical phase diagram of such an open system by studying the feature of the steady-state solutions of Eqs.(\ref{Arr})-(\ref{Bgr}). While we set $\dot \rho_{k,r(g)r}=0$ those equations can be simplified into a pair of coupled stationary equations:
\begin{eqnarray}
	{\left (\Delta_{A,eff}^s\right )^2} &=& \frac{{4{\Omega_{A} ^2}\left( {1 - 2{\rho _{A,rr}^s}} \right) - {\rho _{A,rr}^s}}}{{4{\rho _{A,rr}^s}}} ,\label{DAF} \\
	{\left(\Delta_{B,eff}^s\right)^2} &=& \frac{{4{\Omega_{B} ^2}\left( {1 - 2{\rho _{B,rr}^s}} \right) - {\rho _{B,rr}^s}}}{{4{\rho _{B,rr}^s}}}, \label{DBF}
\end{eqnarray}
where the superscript $s$ denotes the steady-state solutions. $\Delta_{A(B),eff}^s$ is defined as in (\ref{DeltaA}) and (\ref{DeltaB}) by replacing $\rho_{A(B),rr}$ with $\rho_{A(B),rr}^s$.
We find Eqs. (\ref{DAF}) and (\ref{DBF}) can give out nine pairs of roots ($\rho_{A,rr}^s$ and $\rho_{B,rr}^s$), in which only the real ones with values belonging to $ [0,0.5]$ are physical, since for single two-level atom the excitation probability saturates to 0.5 \cite{petrosyan13}. 
The stability of these roots can be tested by adding small perturbations and see whether the system can be eventually settled on these solutions. A detailed description of studying the stability criterion is presented elsewhere, e.g. the supplement of ref. \cite{lee11}.

We classify all the dynamical phases according to the number of the steady-state solutions as well as their dynamical features. As summarized in Table \ref{summ}, the total number of physical roots (PRN) is displayed in the second row, and the number of stable (SRN) and unstable roots (USRN) are in the third row and the fourth row, respectively. If SRN$>$USRN the dynamical phase is a stable phase; oppositely, it is an unstable phase. By given parameters, in the current scheme we find totally three stable as well as five unstable phases. 
Stable phases include the antiferromagnetic phase (labeled by 1AF) with SRN=1 and USRN=0; the bistable antiferromagnetic phase (labeled by 2AF) with SRN=2 and USRN=1; the tristable antiferromagnetic phase (labeled by 3AF) with SRN=3 and USRN=2. Other five unstable phases are, respectively, labeled by CH1, CH2, CH3, CH4  and CH5 whose SRN and USRN can be checked in Table \ref{summ} .In the following discussions, we will show that the dynamics of these stable phases always settles on one of the stable roots for all cases. However, as for the unstable phases it is quite different. In the weak-interaction case where the RRIs are comparable or smaller than the optical coupling strength, the unstable phases tend to show simply oscillatory dynamics; while in the strong-interaction case in which the RRIs play the dominant roles, that is, $U_{AB}\gg\Omega_{A},\Omega_{B}$, the system dynamics easily tends to be chaotic. 
The significant difference from the chaotic dynamics to the ordinary oscillatory dynamics is its continuous frequency spectrum without any characteristic frequency as shown in the inserts of Fig. \ref{phase1} below. Besides, we note that although it is demanding to implement, for the phases CH3, CH4, CH5, if the system is initially prepared properly, very closing to their stable roots, the dynamics may also be stable and without oscillation.

\begin{table}[tbp]
\begin{tabular}{c|ccc|cccccc}
\hline\hline
& \multicolumn{3}{c}{stable phases} & \multicolumn{5}{c}{unstable phases} \\
\hline 
 PRN & 1 & 3 & 5 & 1 & 3 & 3 & 5 & 5 \\ 
SRN & 1 & 2 & 3 & 0 & 0 & 1 & 1 & 2 \\ 
USRN & 0 & 1 & 2 & 1 & 3 & 2 & 4 & 3 \\ 
Label& 1AF  &  2AF  &  3AF  & CH1   & CH2  & CH3  & CH4  &  CH5\\
\hline\hline
\end{tabular}%
\caption{Eight types of possible dynamical phases obtained in the chain of heterospecies Rydberg atoms with the impact of NN and NNN RRIs. PRN represents the total number of physical roots of Eq. (\ref{DAF}) and (\ref{DBF}) (see text), and SRN and USRN stand for the number of the stable and unstable roots, respectively. The phases are stable if SRN is larger than USRN. The three stable phases are labeled by $n$AF while the five unstable phases by CH$n$.}%
\label{summ}
\end{table}
	
In what follows we would discuss the conditions of the emergence of these phases and study their changes with or without the effect of NNN RRIs by numerically simulating the system dynamics.
Before that, we have to make a realistic estimation for all parameters required in the calculation. We use the decay rate $\gamma$ as the frequency unit and assume the condition $\Omega_A=\Omega_B=\Omega$ with $\Omega=10.0$ for strong optical coupling cases and $\Omega=2.0$ for weak coupling cases. Besides, the strong and weak NN RRIs are represented by $U_{AB}=50.0$ and $U_{AB}=10.0$, respectively. The validity of these parameters can be verified by assuming $\gamma=0.1$MHz, the resulting RRIs are $U_{AB}=5.0$MHz or $1.0$MHz by using the lattice spacing $d=3.84\mu$m or $5.0\mu$m. The required vdWs interaction coefficient is $C_6^{(ij)}$=16GHz$\mu$m$^6$, as suggested by ref. \cite{schonleber14}.

%although it may break when many-body collective excitation takes an important role in some regimes \cite{garttner14}. In our scheme two adjustable atomic detunings $\Delta_A $ and $ \Delta_B$ would induce an intrinsic population imbalance from laser pumping between two neighboring lattice sites. The competition of this intrinsic imbalance and RRIs-induced excitation blockade effect can lead to more exotic quantum phases.

\section{Phase diagram without NNN RRIs}
In this section we focus on the case without the impact of NNN RRIs, i.e., $U_{AA}=U_{BB}=0$. In experiment, the NN RRIs can be controlled directly by changing the interatomic distance \cite{beguin13} and the detunings $\Delta_A$ and $\Delta_B$ could be easily adjusted by changing laser frequencies. We first fix $\Delta_A=0$ (atom A is resonantly excited) and tune $\Delta_B$ from -20 to 40 to see the change of steady Rydberg population of different atom species given by the Eqs. (\ref{DAF}) and (\ref{DBF}).

\begin{figure}
\includegraphics[width=3.4in,height=3.9in]{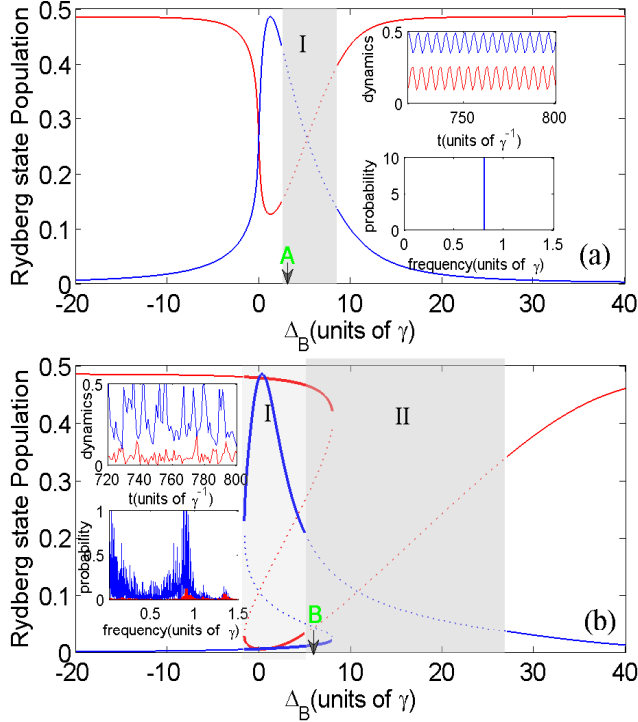}
\caption{(Color online) The stationary solutions $\rho_{A,rr}^s$ (marked by red curves) and $\rho_{B,rr}^s$ (marked by blue curves) as a function of detuning $\Delta_B$ when atoms A are assumed to be resonantly excited. (a) shows the weak interaction case with $U_{AB}=10$ and $\Omega=2$, while (b) for the strong interaction case with $U_{AB}=50$ and $\Omega=2$. Stable and unstable solutions, respectively, are shown by the solid and dotted curves. Insets of (a): the regular oscillatory dynamics and its single-frequency spectrum are observed at $\Delta_B$=3.0 marked by ``A'' in the main figure. Insets of (b): the chaotic dynamics and its continuous frequency spectrum are observed at $\Delta_B$=6.0 marked by ``B'' in the main figure. The initial condition is $\rho_{A,rr}^{t=0}=\rho_{B,rr}^{t=0}=0.1$. Decay rate $\gamma$ is the frequency unit.}
\label{phase1}
\end{figure}

In Figure \ref{phase1} we present the stationary Rydberg state populations with respect to $\Delta_B$, with the weak interaction case ($U_{AB}=10.0$ and $\Omega=2$) in (a) and the strong interaction case ($U_{AB}=50.0$ and $\Omega=2$) in (b). Stable and unstable dynamical phases are, respectively, marked by solid and dotted curves. 
For the weak interaction case when $\Delta_B<0$ or $\Delta_B>8.5$, corresponding to the far off-resonance cases evaluated by the effective detuning $\Delta_{B,eff}^{s}$, 
there merely exists 1AF phase with its dominant Rydberg probability in atom A. 
As $\Delta_B$ grows from negative to small positive values, we find the increase of $\rho_{B,rr}^s$ would make $\Delta^s_{A,eff}$($\propto-U_{AB}\rho^s_{B,rr}$) nonzero, which further yields the excitation probability exchange between atoms A and B. 
However, as $\Delta_B$ increases to the regime \rm{I} where $\Delta_{A,eff}^{s}$ and $\Delta_{B,eff}^{s}$ are comparable, the steady-state solutions are found to become unstable, labeled by the oscillatory phase CH1. In this case the system dynamics can be characterized by the single-frequency oscillations due to the presence of weak Rydberg interactions, as displayed by the inset of (a). 
This finding shows similar results to the single-species atom case in which the dynamics of the system will be periodically oscillating if its corresponding steady-state solutions become unstable (see Fig. 2(b) of reference \cite{lee11}).

We now turn to investigate the strong interaction case in which the Rydberg blockade effect could play a significant role in the system dynamics. Similar to the result of weak-interaction case that  at a negative or large positive $\Delta_B$ the system tends to stay on 1AF phase. We then pay more attention to the center region where $\Delta_{A,eff}^{s}\approx\Delta_{B,eff}^{s}$ is satisfied.  This region can be divided into two parts.

In part \rm{I} there exists only 2AF phase with two stable roots (one is $\rho^s_{A,rr}>\rho^s_{B,rr}$ and the other is $\rho^s_{A,rr}<\rho^s_{B,rr}$) as well as one unstable root. The system will selectively settle on one of the two stable roots accounting for its initial population preparations;

In part \rm{II} the phase transits from CH3 to CH1. CH3, corresponding to one stable and two unstable roots, is an unstable phase. 
Except that initially the system is prepared on a state near the stable root, the system dynamics is oscillatory and trends to be chaotic in the strong interaction limit $U_{AB}\gg\Omega$, characterized by a continuous spectrum in the frequency domain as presented in the insets of (b). As $\Delta_B$ increases, no stable solution is supported by the given parameters and the phase changes into CH1. However, owing to the strong RRIs, this CH1 phase also shows chaotic dynamics instead of the regular oscillations in the weak-interaction cases.
		
%For the single-species atoms, the system may remain chaotic as long as the condition $U_{vdW}\approx \Omega_0$ is met \cite{qian13}. Here $U_{vdW}$ represents the induced energy shifts by strong NN RRIs. Besides, it is worthwhile to point out that once $U_{vdW} \gg \Omega_0$ is satisfied, a perfect blockade would occur between the nearest sites, where only 1AF presents. We are mainly interested in the partial blockade regime where $U_{vdW}\approx\Omega_0$ \cite{barredo14} as a result a large variety of phases would present owing to the competition effect.
		
\begin{figure}
\includegraphics[width=3.4in,height=2.7in]{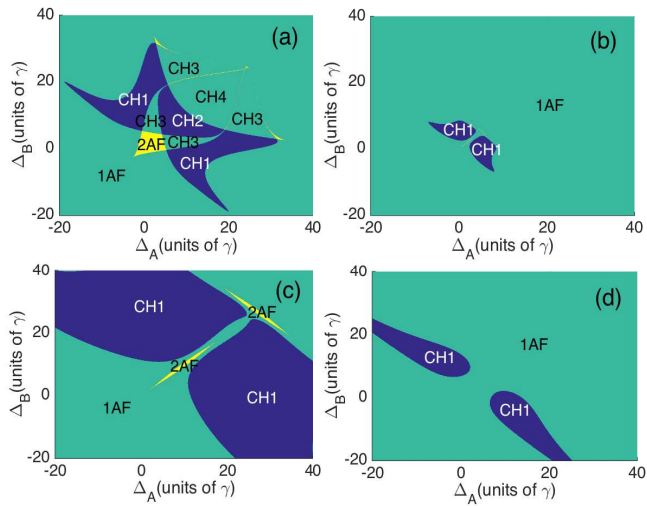}
\caption{(Color online) Phase diagram in the two-dimensional parameter space $\left(\Delta_A,\Delta_B \right )$. The left-column diagrams are obtained from the strong-interaction case with $U_{AB}=50.0$, $\Omega=2.0$ in (a) and $U_{AB}=50.0$, $\Omega=10.0$ in (c). The right-column diagrams are from the weak-interaction case with $U_{AB}=10.0$, $\Omega=2.0$ in (b) and $U_{AB}=10.0$, $\Omega=10.0$ in (d). $\gamma$ is the frequency unit.}
\label{phase2}
\end{figure}
	
In Fig. \ref{phase2} we plot the phase diagrams in a two-dimensional (2D) parameter space of $\Delta_A$ and $\Delta_B$, performing a comparison among the four different cases. In case (b) where the NN interaction strength $U_{AB}=10$ and the optical coupling $\Omega=2$, we find the phase diagram is mainly occupied by the stable phase 1AF, and the unstable phase CH1 only survives in two narrow areas where $\Delta_A$ and $\Delta_B$ have opposite signs. When the optical coupling increases to be comparable with the interaction strength, as in case (d) with $U_{AB}=\Omega=10$, the regions of CH1 expand, but 1AF is still the dominant phase. We stress that due to the small ratio of interaction $U_{AB}$ and $\Omega$, in both cases (b) and (d) the dynamics of unstable phase CH1 is only normal oscillation rather than chaos. When we turn to case with a very strong interaction strength and a weak optical coupling, such as case (a) with $U_{AB}=50$ and $\Omega=2$, besides the phases 1AF and CH1, the phase diagram shows a large number of unique phases, including the bistable phase 2AF and unstable phases CH2, CH3, and CH4. Moreover, we find the dynamics of these unstable phases are no longer regular oscillations, but chaotic oscillations without any characteristic frequencies, which is similar to the case shown in the inset of Fig.\ref{phase1}(b). In case (c) we keep $U_{AB}=50$ and increase $\Omega$ to $10$. Due to narrowing the gap between the interaction strength and the optical coupling, the phase diagram degenerates and resembles the ones for cases (b) and (d). CH1 becomes the only survival unstable phase and 2AF shrinks into narrow areas where the detunings of atoms A and B are almost equal. Also the dynamics of CH1 turns back into normal oscillation, which reconfirms the condition for arising chaos in unstable phases is strong interaction limit $U_{AB}\gg\Omega$.

\section{Phase diagram with considerable NNN RRIs}	
In the following, we consider the effect of NNN RRIs on phase diagram of the system. For simplicity, we assume the long-range NNN RRIs for atoms of different species are equal and is half of the NN RRIs, that is,  $U_{AA}=U_{BB}=U_0=\pm0.5U_{AB}$. Plus and minus represent repulsive and attractive NNN RRIs, respectively.

\begin{figure}
\includegraphics[width=3.4in,height=2.5in]{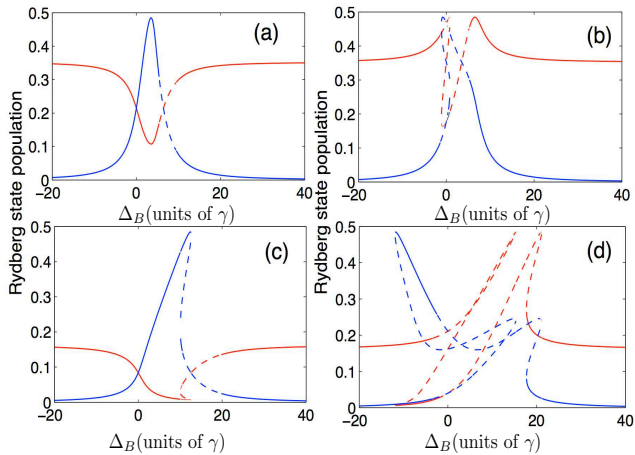}
\caption{(Color online) Stationary Rydberg population $\rho_{A,rr}^s$(red curves) and $\rho_{B,rr}^s$(blue curves) with respect to $\Delta_B$. Stable and unstable solutions are marked by solid and dashed curves, respectively. The long-range NNN interaction is $U_0=0.5U_{AB}$ in (a), (c) and $U_0=-0.5U_{AB}$ in (b), (d). (a)-(b) present the weak-interaction case with $U_{AB}=10.0$ and $\Omega=2.0$, and (c)-(d) present the strong-interaction case with $U_{AB}=50.0$ and $\Omega=2.0$. $\Delta_A$ is fixed at $0$ for all the cases.}
\label{phase3}
\end{figure}

Figure \ref{phase3} shows the change of steady-state Rydberg population $\rho_{A,rr}^s$ and $\rho_{B,rr}^s$ when the RRIs between NNN-site atoms are included. 
(a) and (c) are for the weak NN interaction case ($U_{AB}=10$) and the strong interaction case ($U_{AB}=50$), respectively. In both cases we set the NNN interaction is repulsive ($U_0=0.5U_{AB}$) and the detuning $\Delta_A=0$. The arrangement is similar for (b) and (d) but with an attractive NNN interaction.   
Comparing with Fig. \ref{phase1} which neglects the effect of NNN interactions, we observe a significant decrease of population in atom A (labeled by red curves), particularly at the region of large detuning $|\Delta_B|$; because according to Eq. (\ref{DeltaA}), a repulsive $U_0$ could make the effective detuning $\Delta_{A,eff}^s$ more negative besides the existing negative shift caused by the NN interaction term. In the middle region where $\Delta_B$ is positive but small, we numerically find $\Delta_{A,eff}^s\approx\Delta_{B,eff}^s$, and a repulsive NNN interaction can partially compensate for the detunings (e.g. $\Delta_B$ can be compensated by $U_0\rho_{B,rr}^s$ if $\rho_{A,rr}^s$ is small), which gives rise to a reduction to the number of unstable or multi-stable (bistable or tristable) stationary solutions.

However, while turning to the attractive case ($U_0<0$)
we find it changes significantly. Due to the different signs of the NNN interaction and the NN interaction, the effective steady-state atomic detunings can be rewritten as
\begin{eqnarray}
	{\Delta_{A,eff}^s} &=&{  \Delta_A+(|U_0|\rho_{A,rr}^s-U_{AB}\rho_{B,rr}^s}) , \label{DA}\\
	{\Delta_{B,eff}^s} &=& { \Delta_B+(|U_0|\rho_{B,rr}^s-U_{AB}\rho_{A,rr}^s}). \label{DB}
\end{eqnarray}
Different from the repulsive case, here the frequency shifts caused by the NNN interaction and the NN interaction are opposite. 
As demonstrated by Fig. \ref{phase3}(b) where $U_{AB}=10.0$ is relatively weak, we see $\rho_{A,rr}^s$ could touch the peak value $0.5$ again when the frequency shifts in the effective detuning $\Delta_{A,eff}^s$ caused by the NN interaction and NNN interaction cancel each other.
As $|\Delta_B|$ approaching $0$, $\rho_{A,rr}^s$ becomes unstable and shows a sharp deep, in which a clear population exchange occurs between atom A and atom B. That is because atom B is almost resonantly excited instead of atom A when $\Delta_{B,eff}^s$ is close to zero due to the offset effect of the NN and the NNN interactions in $\Delta_{B,eff}^s$.
With a larger $U_{AB}$ value it will further lead to multiple unstable solutions, as displayed in Fig. \ref{phase3}(d) where the dynamics of system becomes more complex and elusive.

\begin{figure}
\includegraphics[width=3.4in,height=2.75in]{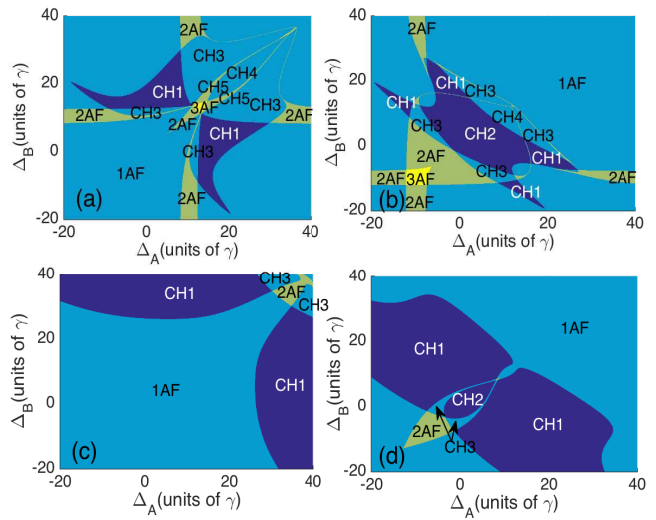}
\caption{(Color online) Phase diagram in $\left (\Delta_A,\Delta_B \right )$ space by the effect of repulsive and attractive NNN RRIs. (a) and (b) correspond to the case of Fig. \ref{phase2}(a) where $U_{AB}=50.0$ and $\Omega=2.0$; (c) and (d) to Fig. \ref{phase2}(c) where $U_{AB}=50.0$ and $\Omega=10.0$. Correspondingly, we use the NNN RRIs $U_0=0.5U_{AB}$ (repulsive) in (a) and (c) and $U_0=-0.5U_{AB}$ (attractive) in (b) and (d).}
\label{phase4}
\end{figure}

Mapping these results into a 2D parameter space of $\Delta_A$ and $\Delta_B$ and compare them with the cases in Fig. \ref{phase2}(a) and (c), we could more clearly find the change of dynamical phases under the impact of the considerable NNN RRIs. As shown in Fig. \ref{phase4},  generally speaking, it depends on the NNN RRIs are repulsive ($U_0>0$) or attractive ($U_0<0$) that all the unique phases except 1AF phase would diffuse from or converge to the center region where $\Delta_{A}\approx\Delta_{B}\approx 0$. 
As we analyzed in the previous paragraph, that is due to the additive effect of the repulsive NNN interaction and the NN interaction on the effective detunings $\Delta_{A,eff}^s$ and $\Delta_{B,eff}^s$. A typical example is Fig. \ref{phase4}(a) whose parameters are same as Fig. \ref{phase2}(a), except for NNN interaction $U_0=0.5$. We find the bistable phase 2AF occurs mainly when $\Delta_{A}$ or $\Delta_{B}$ is far off resonance, in order to compensate for the far difference between $\rho_{A,rr}^s$ and $\rho_{B,rr}^s$. A new tristable phase 3AF arises due to the multiple steady state solutions caused by the including of the NNN RRIs. Other unstable phases are all dispersedly distributed in the parameter space and CH2 phase disappears. In Fig. \ref{phase4}(c) and (d) where both the NN interaction and the optical coupling are large, the number of phases dramatically decreases, but compare with Fig. \ref{phase2}(c) we can still find the similar diffusion and convergence effect caused by the NNN interactions.

Finally, we stress again the properties of all unique phases possibly obtained in our scheme when both the NN and NNN RRIs are considered. A related result is also organized in TABLE \ref{summ}.
The three stable phases are
i) 1AF that includes one stable solution and no unstable solution so that the system finally stays on the stable solution;
ii) 2AF that includes two stable solutions and one unstable solution so that the system tends to stop on the one near its initial state;
iii) 3AF that includes three stable solutions and two unstable solutions so that the system would also choose the nearest one to settle. In addition, the five unstable phases are labeled by CH$n$($n=1,2,3,4,5$) in which USRN is larger than SRN. Except that initially the system is very close to the stable roots of CH$n$ ($n=3,4,5$), the system will keep evolving and not settle on any stationary state. In the strong-interaction case where the RRI plays a dominant role, the system may show complex chaotic dynamics with its dynamics to be very sensitive to the initial prepared population on the Rydberg state; in the weak-interaction case with the RRI comparable to the optical coupling strength, these unstable phases will lead to regular oscillatory dynamics for the system.

\section{Influence of NNN RRIs on the Chaotic phase} \label{chaoDS}

In the strong-interaction case with $U_{AB}\gg \Omega$ we have found the dynamics of the system can show chaotic properties if the dynamical phase becomes unstable.
Due to the high sensitivity of the chaos, it becomes a good candidate to show the dramatic influence of NNN RRIs on our model.
Here, we use CH4 (SRN=1, USRN=4) as an example to see the final state. By directly solving the dynamical evolution of motional equations (\ref{Arr})-(\ref{Bgr}) we could determine which phase the system will finally evolve into. In the calculation, we assume the initial preparations $\rho_{A,rr}^{t=0}$ and $\rho_{B,rr}^{t=0}$ are fully adjustable.

\begin{figure}
\includegraphics[width=3.45in,height=1.4in]{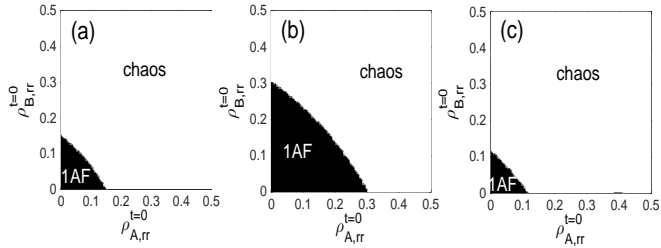}
\caption{Final state distribution in the unstable phase CH4 as functions of $\rho_{A,rr}^{t=0}$ and $\rho_{B,rr}^{t=0}$. (a)-(c), respectively, represent the cases without the NNN interactions (according to Fig. \ref{phase2}(a), $U_0$=0), with repulsive NNN interactions (according to Fig. \ref{phase4}(a), $U_0$=0.5$U_{AB}$) and attractive NNN interactions (according to Fig. \ref{phase4}(b), $U_0$=$-0.5U_{AB}$). The detunings are chosen to be $\Delta_{A}$=$\Delta_B$=15.0, 23.0, 10.0 in (a)-(c).}
\label{Stability}
\end{figure}

Figure \ref{Stability}(a)-(c) display the final state of the system if the initial preparations on the Rydberg states are varied. Since CH4 contains one stable solution that the system may also possibly stay on it when the initial prepared population is close to that stable solution. Thus, when the system settles on CH4 its real dynamics can have two different cases. One is a stable steady state like 1AF (denote by 1AF) and the other is unstable chaos (denoted by chaos).
From (a)-(c) we respectively consider the cases without and with repulsive and attractive Rydberg interactions between two NNN-site atoms. We find, comparing to (a), a repulsive NNN interaction (see (b)) would make the system more stable against chaos. It is more likely to have a stable dynamics under the environment of the chaotic phase CH4. However, when the NNN interaction is attractive as plotted in (c), clearly we find the area in the parameter space for presenting stable dynamics deeply shrinks. Except when $\rho_{A,rr}^{t=0}$ and $\rho_{B,rr}^{t=0}$ are both close to the exact steady state solutions [$\rho_{A,rr}^s=\rho_{B,rr}^s=0.046$], the dynamics of system will be totally chaotic and unable to be measured.
This finding has been implied in Fig. \ref{phase3}(d) in which multiple unstable roots are represented due to $U_0 < 0$.

\section{Conclusions}

We represent a rich variety of dynamical phases of a chain of two-species Rydberg atoms held in a 1D optical lattice, where the optical transition between the atomic ground state and the high-lying Rydberg state is performed by one-photon excitation. 
In particular, except for the NN Rydberg interaction between atoms of different species, the long-range NNN interactions between atoms of same species are also considered. We show that the phase diagrams change a lot when the long-range interaction is included. Especially, the repulsive or attractive long-range interactions can give rise to a clear diffusing or converging effect on the original phase diagrams without them. For instance, a repulsive NNN RRI can help to stabilize the system against chaotic dynamics. In addition, we also study the real final state of the chaotic phase in the strong-interaction limit and display its sensitivity to the initial atomic preparations on the Rydberg state.
Most of these results and phenomena are novel, but it is worthwhile to stress when $\Delta_A = \Delta_B$ and the NNN RRI vanishes, our results will trend to be consistent with the previous findings from the Rydberg system of single-species atoms \cite{lee11}.

This work was supported by the NSFC under Grants No. 11474094, No. 11104076, the
Specialized Research Fund for the Doctoral Program of Higher Education No.
20110076120004.


\bigskip

\begin{thebibliography}{99}

\bibitem{jaksch98} D. Jaksch, C. Bruder, J. I. Cirac, C. W. Gardiner, and P. Zoller, Phys. Rev. Letts. \textbf{81} 3108 (1998).

\bibitem{bloch05} I. Bloch, Nat. Phys. \textbf{1} 23 (2005).

\bibitem{morsch06} O. Morsch and M. Oberthaler, Rev. Mod. Phys. \textbf{78} 179 (2006).

\bibitem{lewenstein07} M. Lewenstein, A. Sanpera, V. Ahufinger, B. Damski,
A. Sen De, and U. Sen, Adv. Phys. \textbf{56} 243 (2007)

\bibitem{lewenstein12} M. Lewenstein, A. Sanpera, and Veronica Ahufinger, {\it Ultracold atoms in optical lattices: Simulating quantum many-body systems} (Oxford University Press, 2012).

\bibitem{bloch08} I. Bloch, J. Dalibard and W. Zwerger, Rev. Mod. Phys. \textbf{80} 885 (2008) and references therein.

\bibitem{anderson11} S. E. Anderson, K. C. Younge, and G. Raithel, Phys. Rev. Letts. \textbf{107} 263001 (2011).

\bibitem{gallagher08} T. F. Gallagher and P. Pillet, Adv. At. Mol. Opt. Phys., \textbf{56} 161 (2008).

\bibitem{comparat10} D. Comparat and P. Pillet, JOSAB \textbf{27}, A208 (2010).

\bibitem{pohl10} T. Pohl, E. Demler and M. D. Lukin, Phys. Rev. Letts. \textbf{104}, 043002 (2010).

\bibitem{weimer10} H. Weimer, M. M\"{u}ller, I. Lesanovsky, P. Zoller and H. P. B\"{u}chler, Nat. Phys. \textbf{6} 382 (2010).

\bibitem{viteau11} M. Viteau, M. G. Bason, J. Radogostowicz, N. Malossi, D. Ciampini, O. Morsch, and E. Arimondo, Phys. Rev. Letts. \textbf{107} 060402 (2011).

\bibitem{schaub12} P. Schaub, M. Cheneau, M. Endres, T. Fukuhara, S. Hild, A. Omran, T. Pohl, C. Gross, S. Kuhr and I. Bloch, nature \textbf{491} 87 (2012).


\bibitem{carr13} C. Carr, R. Ritter, C. Wade, C. Adams, and K. Weatherill, Phys. Rev. Letts. \textbf{111}, 113901 (2013).

\bibitem{lee11} Tony E. Lee, H. H\"{a}ffner and M. C. Cross, Phys. Rev. A, \textbf{84} 031402 (2011); I. Lesanovsky, Phys. \textbf{4}, 71 (2011).

\bibitem{qian12} J. Qian, G. Dong, L. Zhou and W. Zhang, Phys. Rev. A, \textbf{85} 065401 (2012).

\bibitem{qian13} J. Qian, L. Zhou and W. Zhang, Phys. Rev. A, \textbf{87} 063421 (2013).

\bibitem{hoening14} M. Hoening, W. Abdussalam, M. Fleischhauer, and T. Pohl, Phys. Rev. A, \textbf{90} 021603 (2014).


\bibitem{olmos14} B. Olmos, D. Yu, and I. Lesanovsky, Phys. Rev. A \textbf{89}, 023616 (2014).

\bibitem{chan15} C-K Chan, T. E. Lee and S. Gopalakrishnan, Phys. Rev. A, \textbf{91} 051601(R) (2015).

\bibitem{honing13} M. H\"{o}ning, D. Muth, D. Petrosyan, and M. Fleischhauer, Phys. Rev. A, \textbf{87} 023401 (2013).

\bibitem{hsueh13} C. Hsueh, Y. Tsai, K. Wu, M. Chang and W. Wu, Phys. Rev. A, \textbf{88} 043646 (2013).

\bibitem{teixeira15} R. C. Teixeira, C. H-Avigliano, T. L. Nguyen, T. C-Moltrecht, J. M. Raimond, S. Haroche, S. Gleyzes and M. Brune, Phys. Rev. Letts., \textbf{115} 013001 (2015).

\bibitem{barredo15} D. Barredo, H. Labuhn, S. Ravets, T. Lahaye, and A. Browaeys, Phys. Rev. Letts. \textbf{114} 113002 (2015).

\bibitem{scelle13} R. Scelle, T. Rentrop, A. Trautmann, T. Schuster, and M. K. Oberthaler, Phys. Rev. Lett. \textbf{111}, 070401 (2013).

\bibitem{ramos14} T. Ramos, H. Pichler, A. J. Daley, and P. Zoller, Phys. Rev. Letts. \textbf{113}, 237203 (2014).

\bibitem{singer05} K. Singer, J. Stanojevic, M. Weidem\"{u}ller and R. C\^{o}t\'{e}, J. Phys. B, \textbf{38} s295 (2005).

\bibitem{beguin13} L. B\'{e}guin, A. Vernier, R. Chicireanu, T. Lahaye, and A. Browaeys, Phys. Rev. Letts. \textbf{110} 263201 (2013).

\bibitem{olmos11} B. Olmos, W. Li, S. Hofferberth, and I. Lesanovsky, Phys. Rev. A, \textbf{84} 041607 (2011).

\bibitem{zoubi15} H. Zoubi, Arxiv: 1507.04114.

\bibitem{diehl08} S. Diehl, A. Micheli, A. Kantian B. Kraus, H.P. B\"{u}chler, and P. Zoller, Nat. Phys. \textbf{4} 878 (2008).

\bibitem{diehl10} S. Diehl, A. Tomadin, A. Micheli, R. Fazio, and P. Zoller, Phys. Rev. Letts. \textbf{105} 015702 (2010).

\bibitem{tomadin11} A. Tomadin, S. Diehl, and P. Zoller, Phys. Rev. A, \textbf{83} 013611 (2011).

\bibitem{ates07} C. Ates, T. Pohl, T. Pattard and J. M. Rost, Phys. Rev. Letts. \textbf{98}, 023002 (2007).

\bibitem{amthor10} T. Amthor, C. Giese, C. S. Hofmann, and M. Weidem\"{u}ller, Phys. Rev. Letts. \textbf{104}, 013001 (2010).

\bibitem{petrosyan13} D. Petrosyan, Phys. Rev. A, \textbf{87} 053414 (2013).

\bibitem{schonleber14} D. W. Sch\"{o}nleber, M. G\"{a}rttner, and J. Evers, Phys. Rev. A, \textbf{89} 033421 (2014).

\bibitem{barredo14} D. Barredo, S. Ravets, H. Labuhn, L. B\'{e}guin, A. Vernier, F. Nogrette, T. Lahaye, and A. Browaeys, Phys. Rev. Letts. \textbf{112} 183002 (2014).

%%%%%%%%%%%%%%%%%%%%%%%%%%%%%%%%%%%%%%%%%%%%%%%%%%%%%




%\bibitem{chin10} C. Chin, R. Grimm, P. Julienne, and E. Tiesinga, Rev. Mod. Phys., \textbf{82} 1225 (2010).


%\bibitem{forges15} L. de Forges de Parny, V. G. Rousseau and T. Roscilde, Phys. Rev. Letts. \textbf{114} 195302 (2015).





%\bibitem{greiner02} M. Greiner, O. Mandel, T. Esslinger, T. W. H\"{a}nsch and I. Bloch, Nature \textbf{415} 39 (2002).


%\bibitem{xu14} Z. Xu, W. S. Cole, and S. Zhang, Phys. Rev. A, \textbf{89} 051604(R) (2014).

%\bibitem{bhaseen09} M. Bhaseen, A. Silver, M. Hohenadler, and B. Simons, Phys. Rev. Letts. \textbf{103} 265302 (2009).

%\bibitem{mishra07} T. Mishra, R. V. Pai, and B. P. Das, Phys. Rev. A, \textbf{76} 013604 (2007).

%\bibitem{papp08} S. Papp, J. Pino, and C. Wieman, Phys. Rev. Letts. \textbf{101}, 040402 (2008).

%\bibitem{zhou08} L. Zhou, J. Qian, H. Pu, W. Zhang and H. Ling, Phys. Rev. A, \textbf{78} 053612 (2008).

%\bibitem{lv14} J. Lv, Q. Chen and Y. Deng, Phys. Rev. A, \textbf{89} 013628 (2014).





































%%%%%%%%%%%%%%%%%%






%\bibitem{hart15} R. Hart, P. Duarte, T. Yang, X. Liu, T. Paiva, E. Khatami, R. Scalettar, N. Trivedi, D. Huse and R. Hulet, Nature \textbf{519} 211 (2015).










%\bibitem{catani08} J. Catani, L. Sarlo, G. Barontini, F. Minardi, and M. Inguscio, Phys. Rev. A, \textbf{77} 011603 (2008).









%\bibitem{ji14} S. Ji, V. Sanghai, C. Ates, and I. Lesanovsky, Phys. Rev. A., \textbf{89} 021404(R) (2014).

%\bibitem{gallagher94} T. F. Gallagher, {\it Rydberg atoms} (Cambridge University, Cambridge, England, 1994).



%\bibitem{saffman08} M. Saffman and K. M\o lmer, Phys. Rev. A, \textbf{78} 012336 (2008).

%\bibitem{buluta09} I. Buluta and F. Nori, Science \textbf{326} 108 (2009).



%\bibitem{saffman10} M. Saffman, T. G. Walker and K. M\"{o}lmer, Rev. Mod. Phys. \textbf{82} 2313 (2010).





%\bibitem{malossi14} N. Malossi, M. M. Valado, S. Scotto, P. Huillery, P. Pillet, D. Ciampini, E. Arimondo, and O. Morsch, Phys. Rev. Lett. \textbf{113}, 023006 (2014).



%\bibitem{marcuzzi14} M. Marcuzzi, E. Levi, S. Diehl, J. P. Garrahan, and I. Lesanovsky, Phys. Rev. Lett. \textbf{113}, 210401 (2014).
















%\bibitem{baluktsian13} T. Baluktsian, B. Huber, R. Löw, and T. Pfau, Phys. Rev. Lett. \textbf{110}, 123001 (2013).

%\bibitem{glatzle15} A. W. Gl\"{a}tzle, M. Dalmonte, R. Nath, C. Gross, I. Bloch and P. Zoller, Phys. Rev. Lett. \textbf{114}, 173002 (2015).

%\bibitem{glaetzle14} A. W. Glaetzle, M. Dalmonte, R. Nath, I. Rousochatzakis, R. Moessner, and P. Zoller, Phys. Rev. X \textbf{4}, 041037 (2014).

%\bibitem{wuster13} S. W\"{u}ster, S. M\"{o}bius, M. Genkin, A. Eisfeld, and J. M. Rost, Phys. Rev. A, \textbf{88} 063644 (2013).

%\bibitem{henkel10} N. Henkel, R. Nath, and T. Pohl, Phys. Rev. Lett. \textbf{104}, 195302 (2010).



%\bibitem{levi15} E. Levi, J. Min\'{a}\^{r}, J. P. Garrahan and I. Lesanovsky, arXiv:1503.03259.









%\bibitem{li13} W. Li, C. Ates, and I. Lesanovsky, Phys. Rev. Letts. \textbf{110} 213005 (2013).

%\bibitem{bergmann98} K. Bergmann, H. Theuer and B. W. Shore, Rev. Mod. Phys. \textbf{70} 1003 (1998).





%\bibitem{tannoudji92} C. Cohen-Tannoudji, J. Dupont-Roc, and G. Grynberg, {\it Atom- Photon Interactions} (Wiley, New York, 1992).


%%%%%%%%%%%%%%%%%%%%%%%%%%%%%%%%%%%%%%%%%%%%%%%%%
































%\bibitem{hu13} A. Hu, T. E. Lee, and C. W. Clark, Phys. Rev. A \textbf{88}, 053627 (2013).
















%\bibitem{carmichael93} H. J. Carmichael, {\it An Open Systems Approach to Quantum Optics} (Springer, Berlin, Heidelberg, 1993).

%\bibitem{dalibard92} J. Dalibard, Y. Castin, and K. M\o lmer, Phys. Rev. Lett. \textbf{68}, 580 (1992).

%\bibitem{plenio98} M. B. Plenio and P. L. Knight, Rev. Mod. Phys. \textbf{70} 101 (1998).

%\bibitem{osychenko11} O. N. Osychenko, G. E. Astrakharchik, Y. Lutsyshyn, Yu. E. Lozovik, and J. Boronat, Phys. Rev. A \textbf{84}, 063621 (2011).













%\bibitem{galam98} S. Galam, C. S. O. Yokoi and S. R. Salinas, Phys. Rev. B., \textbf{57} 8370 (1998).

%%%%%
 





















%\bibitem{ditzhuijzen08} C. S. E. van Ditzhuijzen, A. F. Koenderink, J. V. Hern\'{a}ndez, F. Robicheaux, L. D. Noordam, and H. B. van Linden van den Heuvell, Phys. Rev. Letts. \textbf{100} 243201 (2008).
%
%\bibitem{gunter13} G. G\"{u}nter, H. Schempp, M. Robert-de-Saint-Vincent, V. Gavryusev, S. Helmrich, C. S. Hofmann, S. Whitlock and M. Weidem\"{u}ller, Science \textbf{342} 954 (2013).
%
%\bibitem{li14} W. Li, D. Viscor, S. Hofferberth and I. Lesanovsky, Phys. Rev. Letts. \textbf{112} 243601 (2014).
%
%
%\bibitem{bettelli13} S. Bettelli, D. Maxwell, T. Fernholz, C. S. Adams, I. Lesanovsky and C. Ates, Phys. Rev. A, \textbf{88} 043436 (2013).
%
%\bibitem{maxwell14} D. Maxwell, D. J. Szwer, D. Paredes-Barato, H. Busche, J. D. Pritchard, A. Gauguet, M. P. A. Jones, and C. S. Adams, Phys. Rev. A, \textbf{89} 043827 (2014).
%
%
%
%\bibitem{cinti10} F. Cinti, P. Jain, M. Boninsegni, A. Micheli, P. Zoller, and G. Pupillo, Phys. Rev. Letts. \textbf{105} 135301 (2010).
%
%\bibitem{glaetzle12} A. W. Glaetzle, R. Nath, B. Zhao, G. Pupillo, and P. Zoller, Phys. Rev. A, \textbf{86} 043403 (2012). 
%
%
%
%
%
%
%
%
%
%%\bibitem{bijnen14} R. M. W. van Bijnen and T. Pohl, arXiv:1411.3118.
%
%
%
%
%\bibitem{yan15} D. Yan, Z. Wang, C. Ren, H. Gao, Y. Li and J. Wu, Phys. Rev. A, \textbf{91} 023813 (2015).
%
%
%
%
%
%
%
%
%
%\bibitem{garttner14} M. G\"{a}rttner, S. Whitlock, D. W. Sch\"{o}nleber, and J\"{o}rg Evers, Phys. Rev. Letts. \textbf{113} 233002 (2014).




























\end{thebibliography}
\end{document}